\documentstyle{ioplppt}         
\input epsf
\begin{document}
\title{Electronic structure of VO$_2$ studied by x-ray
photoelectron and x-ray emission spectroscopies}
\author{E~Z~Kurmaev\dag, V~M~Cherkashenko\dag, Yu~M~Yarmoshenko\dag,
St~Bartkowski\ddag, A~V~Postnikov\ddag, M~Neumann\ddag,
L-C~Duda\S, J~H~Guo\S, J~Nordgren\S,
V~A~Perelyaev\P\ and W~Reichelt$^*$}
\address{\dag\ Institute of Metal Physics, Russian Academy of Sciences --
Ural Division, Yekaterinburg GSP-170, Russia}
\address{\ddag\ Universit\"at Osnabr\"uck -- Fachbereich Physik,
D-49069 Osnabr\"uck, Germany}
\address{\S\ Physics Department, Uppsala University, Box 530,
S-75121 Uppsala, Sweden}
\address{\P\ Institute of Solid State Chemistry,
Russian Academy of Sciences -- Ural Division, Yekaterinburg GSP-145, Russia}
\address{$\!^*$ Technische Universit\"at Dresden,
Institut f\"ur Anorganische Chemie, Mommsenstr. 13,
D-01062 Dresden, Germany}

\begin{abstract}
A VO$_2$ single-crystal has been subject of a combined investigation by
high resolution x-ray photoelectron spectroscopy (XPS), x-ray emission
spectroscopy (XES) with both electron and energy-selective x-ray excitation
(V$L\alpha$-, V$K\beta_5$- and O$K\alpha$-emission) and x-ray absorption
spectroscopy (XAS) (O$1s$).  We performed first principles
tight-binding LMTO band structure calculations of VO$_2$ in both monoclinic
and tetragonal rutile structures and compare the densities of states (DOS)
with the experimental data. From this we conclude that the electronic
structure of VO$_2$ is more bandlike than correlated.
\end{abstract}

\maketitle

\section{Introduction}
\label{sec:intro}

VO$_2$ belongs to transition metal compounds which
exhibit metal-insulator transitions \cite{1}. At $T$=340 K,
VO$_2$ undergoes a phase transition from a semiconductor
with monoclinic structure to a metal with the tetragonal
rutile structure. The nature of the ground-state semiconducting
phase is still rather uncertain. In the semiconducting phase,
the V atoms are paired which was a reason for the
suggestion that the electron-phonon interaction is responsible for the
splitting of the $d$-band and the opening of a band gap \cite{2}.
This idea is supported by band structure calculations according to which
the crystallographic phase transition can be explained by the formation of a
charge-density wave accompanied by a lattice distortion and a subsequent
condensation of phonons \cite{3}. According to Ref.~\cite{4},
there is enough energy gain to account for the metal-insulator transition
through structural distortions that permits a strengthening of
the vanadium $d$--$d$ bonds and a reorganization of the states
near the Fermi level.

On the other hand, some calculations indicate that a crystallographic
distortion is not sufficient to open up a gap, and that the
electron-correlation effects play an important role in the
transition \cite{5}. It was concluded in Ref.~\cite{6}
that the energy gap in VO$_2$ is of the charge-transfer type rather than of the
Mott--Hubbard type as in the late transition metal compounds, like NiO and
CuO \cite{7}.

In addition to earlier spectroscopic studies \cite{13,14,15,16},
the photoelectron spectroscopy measurements and low-energy electron
diffraction studies on VO$_2$ have been carried out recently
in Ref.~\cite{pes,phB}.
The present work aims at a combined experimental study of the
electronic structure of a VO$_2$ single crystal at room temperature by
the use of high-energy spectroscopies. A high-resolution x-ray
photoelectron spectroscopy (XPS) provides information about the total
density of states (DOS) in the valence band (VB); V$L\alpha$
(the $3d4s\rightarrow2p$ transition), V$K\beta_5$ (the $4p\rightarrow1s$
transition), O$K\alpha$ (the $2p\rightarrow1s$ transition) x-ray emission
(XES) VB spectra (excited by both electrons and photons)
probe the V$3d$, V$4p$ and O$2p$ partial DOS in the valence band;
the O$1s$ total electron yield spectrum probes the O$2p$ unoccupied states.
Band structure calculations are performed and compared with the experimental
spectra leading to the conclusion that the electronic structure of VO$_2$ is
more bandlike than correlated.

\section{Experimental procedure}
\label{sec:exp}

The XPS measurements have been carried out with a PHI 5600 ci
multitechnique spectrometer using monochromatized aluminium $K\alpha$
radiation with a full width at half-maximum (FWHM) of 0.3 eV.  The
energy resolution of the analyzer was 1.5\% of the pass energy.  We
estimate an energy resolution of about 0.35 eV for the XPS measurements on
VO$_2$.  The base pressure in the vacuum chamber during the measurements
was $5\times10^{-9}$ Torr.  All of the experiments
presented in this paper have
been performed at room temperature with the same single crystal of VO$_2$
($2\times7\times0.5$ mm).

Initial measurements were performed without
cleaving the crystal, and hence a high contamination
of carbon was detected on the
surface. Therefore the final measurements were
done on a surface that was cleaved {\it in vacuo}. Thus an excellent
surface with a relatively small amount of defects and contaminations could
be obtained, and hence the intrinsic
properties of the samples could be studied.  For the comparison, XPS
measurements of pure V metal (single crystal) were also performed.  All XPS
spectra were calibrated using the Au $4f_{7/2}$ signal from an Au foil
[$E_{b.e.}(4f_{7/2})$ = 84.0 eV].

Electron excited V$L\alpha$-emission spectra (the $3d4s\rightarrow2p_{3/2}$
transition) of VO$_{2}$ were recorded using an RSM-500 spectrometer with a
diffraction grating ($N$=600 lines/mm; $R$=6 m).  The accelerating
voltage and current on the x-ray tube were $V$=4.4 keV and $I$=0.3 mA.
The energy resolution was 0.4 eV.

V$K\beta_{5}$-spectra (the $4p\rightarrow1s$ transition)
of VO$_2$ were measured using a fluorescent Johan-type
vacuum spectrometer with a position-sensitive detector \cite{8}.
Cu$K\alpha$ x-ray radiation from the sealed x-ray
tube was used for excitation for the fluorescent V$K\beta_{5}$ XES.
A quartz crystal (rhombohedral plane, second order reflection)
curved to $R$=1.8 m was used as an analyzer.
The spectra were measured with an energy resolution ${\Delta}E$=0.22 eV.

Energy-selective excited O$K\alpha$- (the $2p\rightarrow1s$ transition),
V$L\alpha$-spectra (the $3d4s\rightarrow2p_{3/2}$ transition) were measured as
well as x-ray absorption spectra (the V$2p$- and O$1s$-edges) in the
sample drain-current mode. These measurements were performed at the
undulator beam line BW3 at HASYLAB Hamburg, Germany \cite{9}, equipped with a
modified SX-700 monochromator.  The soft x-ray emission spectra were
recorded at various excitation energies in the first order of diffraction
with a resolution of about 0.7 eV.  We used a grazing-incidence grating
spectrometer \cite{10} with an $R$=5 m spherical grating with 1200 lines/mm
in a Rowland circle geometry.
The resolution of the excitation radiation was set to about 1 eV by opening
the exit slit of the monochromator to 400 $\mu$m.  The spectrometer had
solid angle acceptance of about $2\times 10^{-5}$ sr, so that with a spot
size of some
$500\times 500$ $\mu$m at a distance of 5 cm from the entrance slit of
the spectrometer we registered a maximum of about 300 counts/minute in the
V$L\alpha$ line at 100 mA ring current. In order to obtain a
reasonable statistical accuracy, we had to acquire for 120 to 240 minutes
per single spectrum, depending on the excitation energy.  The angle between
the incident x-ray beam and the detection direction was 90$^{\circ}$.  This
minimized the elastic scattering into the spectrometer, because the
polarization of the beam coincided with the detection direction.
The V$2p$ and
O$1s$ absorption spectra had energy resolution of 0.2 eV for an exit slit
width of 80 $\mu$m.  No special surface preparation was undertaken for these
measurements.

\begin{figure}[tbh]
\epsfxsize=12.0cm
\centerline{\epsfbox{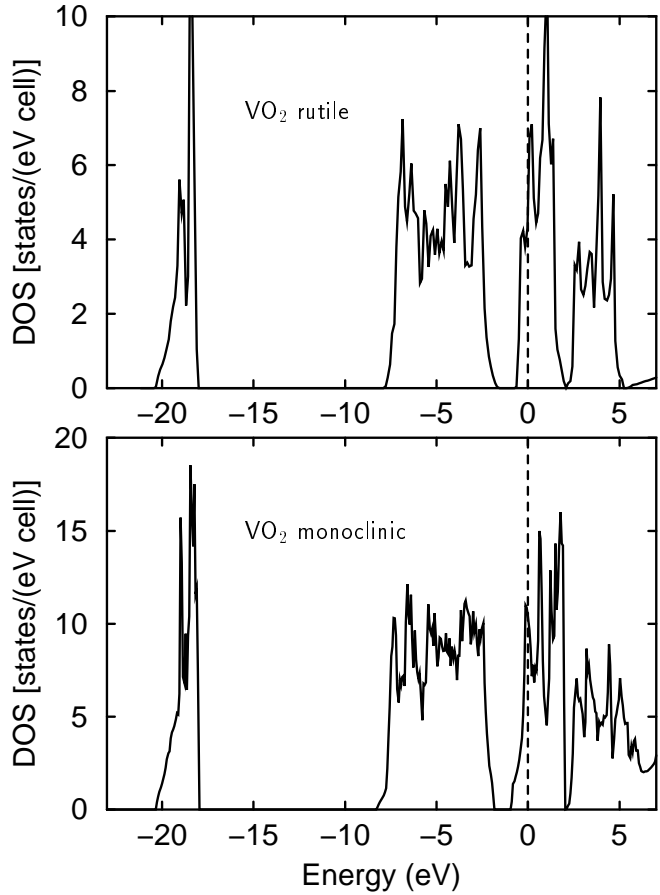}}
\caption{
Total density of states calculated for rutile (tetragonal)
and monoclinic crystal structures of VO$_2$.}
\label{fig_1}
\end{figure}

\section{Results and Discussion}
\label{sec:results}

\subsection{Electronic Structure Calculations}
\label{sec:calc}

The band-structure calculations were carried out by
the tight-binding linear muffin-tin orbital method (TB-LMTO) in the atomic
sphere approximation (ASA) \cite{OKA}, with the use of the
exchange-correlation potential as parametrized by
von Barth and Hedin \cite{vBH} and gradient
corrections as proposed by Langreth and Mehl \cite{LM}
on top of the local density approximation (LDA).
Since both rutile and monoclinic structures of VO$_2$ are relatively
loosely packed, it was necessary to include empty spheres
in the ASA calculations. For the tetragonal rutile phase,
we used the crystal structure parameters $a$=4.5546\AA,
$c/a$=0.626 and $u$=0.300 as given in Ref.~\cite{ru}.
Besides two formula units of VO$_2$, our unit cell
included eight equivalent empty spheres, which form long chains
parallel to the linear O-V-O fragments, separating the latter.
Due to the difference between the V and O atomic sphere radii, the chains
are slightly bent. Our choice of the sphere radii $S$, based on the
attainability of good matching between the potential at the V and O atomic
spheres compatible with a radial sphere overlap below 25\%
of all interatomic distances, was: $S$(V)=2.702 a.u.;
$S$(O)=1.762 a.u.; $S$(empty spheres)=1.619 a.u.
The calculated band structure and the total
DOS are rather close to those obtained earlier
in Ref.~\cite{3,ckz,12} by different methods. We
performed our own calculation because no data
on the partial DOS, that are necessary for the discussion
of the XES, were available from the previous calculations.
Our calculated total DOS for the rutile phase of VO$_2$
is shown in Fig.~\ref{fig_1}, upper panel.

As is usual in oxides, the valence band is formed by
hybridized transition metal $3d$ and O$2p$ states.
The band gap of 0.9 eV within the valence band separates the regions
where O$2p$ states (below the gap) and V$3d$ states (above the gap)
dominate. The Fermi level crosses the
upper subband, revealing a metallic behaviour of VO$_2$,
as consistent with the results of other calculations
done for the rutile-type structure \cite{3,4,ckz,12}.
The O$2s$-related subband (which experiences some hybridization
with the V$3d$ and V$4p$ states) lies separately at about 20 eV
below the Fermi level.
The overall shape of the band structure and of the total DOS
obtained in our calculation is in agreement with
earlier results \cite{3,ckz,12}; the width of the gap
within the valence band is somewhat larger than 0.62 eV
as obtained in the full potential linear augmented plane
wave calculation of Ref.~\cite{12}. The gap width of 4.6 eV
reported in Ref.~\cite{3} seems to be too large (probably
due to the use of the Slater exchange potential) and not
in agreement with the experimental positioning of
individual subbands, as discussed in Ref.~\cite{ckz}
or found in the present paper.

\begin{figure}[tbh]
\epsfxsize=12.5cm
\centerline{\epsfbox{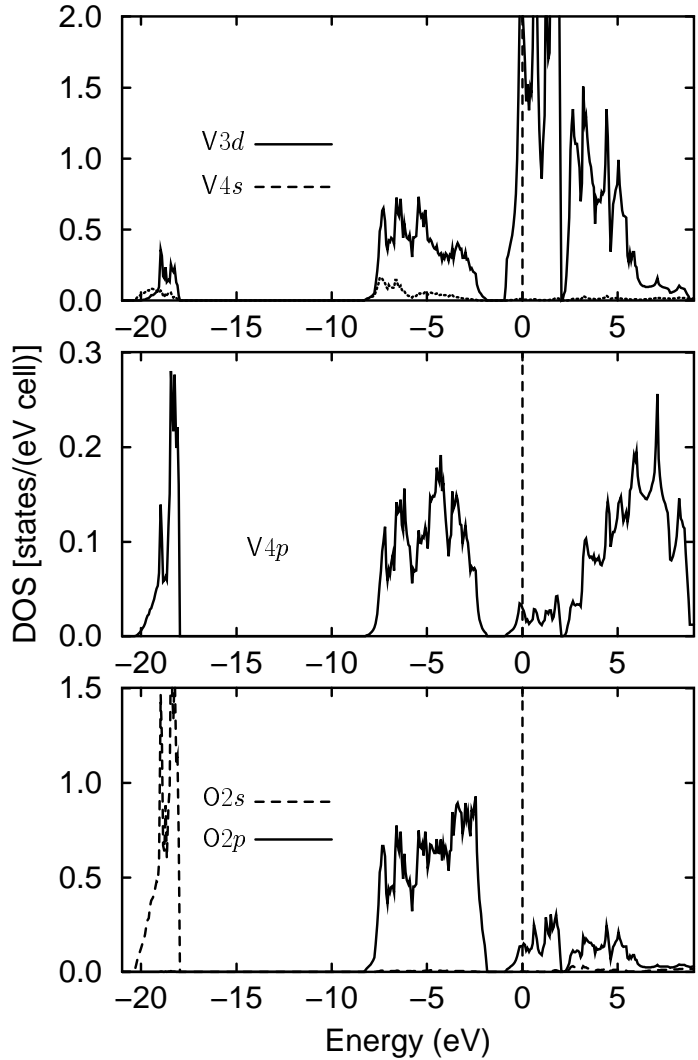}}
\caption{
Partial density of states distribution
in the monoclinic structure of VO$_2$.}
\label{fig_2}
\end{figure}

Since the room-temperature phase (for $T<68^{\circ}$C) of VO$_2$ is
monoclinic, the relevant comparison with experiment needs the calculation
data for the latter structure.  The monoclinic unit cell contains four
formula units of VO$_2$, and, because of the large number of atoms and low
symmetry, only few calculations have been done by now.  An earlier
non-self-consistent calculation \cite{ck} reproduces an experimentally
observable semiconductor band gap, but otherwise seems to be very
inaccurate in describing the overall structure of the valence band.
The analysis of the structure transformation by means of {\it ab~initio}
molecular dynamics in Ref.~\cite{4} shows that the monoclinic phase
has lower energy than the rutile phase,
and the equilibrium positions of the atoms are in good agreement with
the experimental determination. However, the strength of
the tendency for opening
a gap within the valence band is underestimated within the LDA,
and the flat bands in the vicinity of the Fermi level do not
separate completely (see, e.g., Fig. 3 of Ref.~\cite{4}).
This is also the case in our calculation,
which we have carried out for the experimental monoclinic structure
as specified in Ref.~\cite{mon}.
The change from rutile to monoclinic structure gives rise to
a broadening of O$2p$ and V$3d$ bands by $\sim$0.2 eV and to
more pronounced splitting in the V$3d$ states of $t_{2g}$ symmetry.
In Ref.\cite{ni} it was shown
that the inclusion of electron-phonon interaction (in the periodic
shell model, based on the results of discrete-variational X$_{\alpha}$
cluster calculation) leads to the opening of the band gap.
According to our calculation, the distortion of the nested bands
due to the displacement of the atoms in the doubled-cell monoclinic structure
gives rise to only small changes in the partial DOS, as compared
to the rutile structure. Technically, the calculation deals
with two inequivalent
oxygen species and three inequivalent types of lattice-packing empty
spheres of different sizes making a total of 24 sites.  Our calculated
total DOS for the monoclinic structure is shown in Fig.~\ref{fig_1}, lower
panel, and some partial DOS are shown in Fig.~\ref{fig_2}. For comparison
with XES, we did not distinguish the data for two oxygen species and show
the averaged O$2s$ and O$2p$ DOS over all sites in Fig.~\ref{fig_2},
lowest panel. The densities of states are plotted for the
$16\!\times\!16\!\times\!16$ mesh over the Brillouin zone.

\subsection{ X-ray Photoelectron Spectra}
\label{sec:xps}

\begin{figure}[tbh]
\epsfxsize=12.0cm
\centerline{\epsfbox{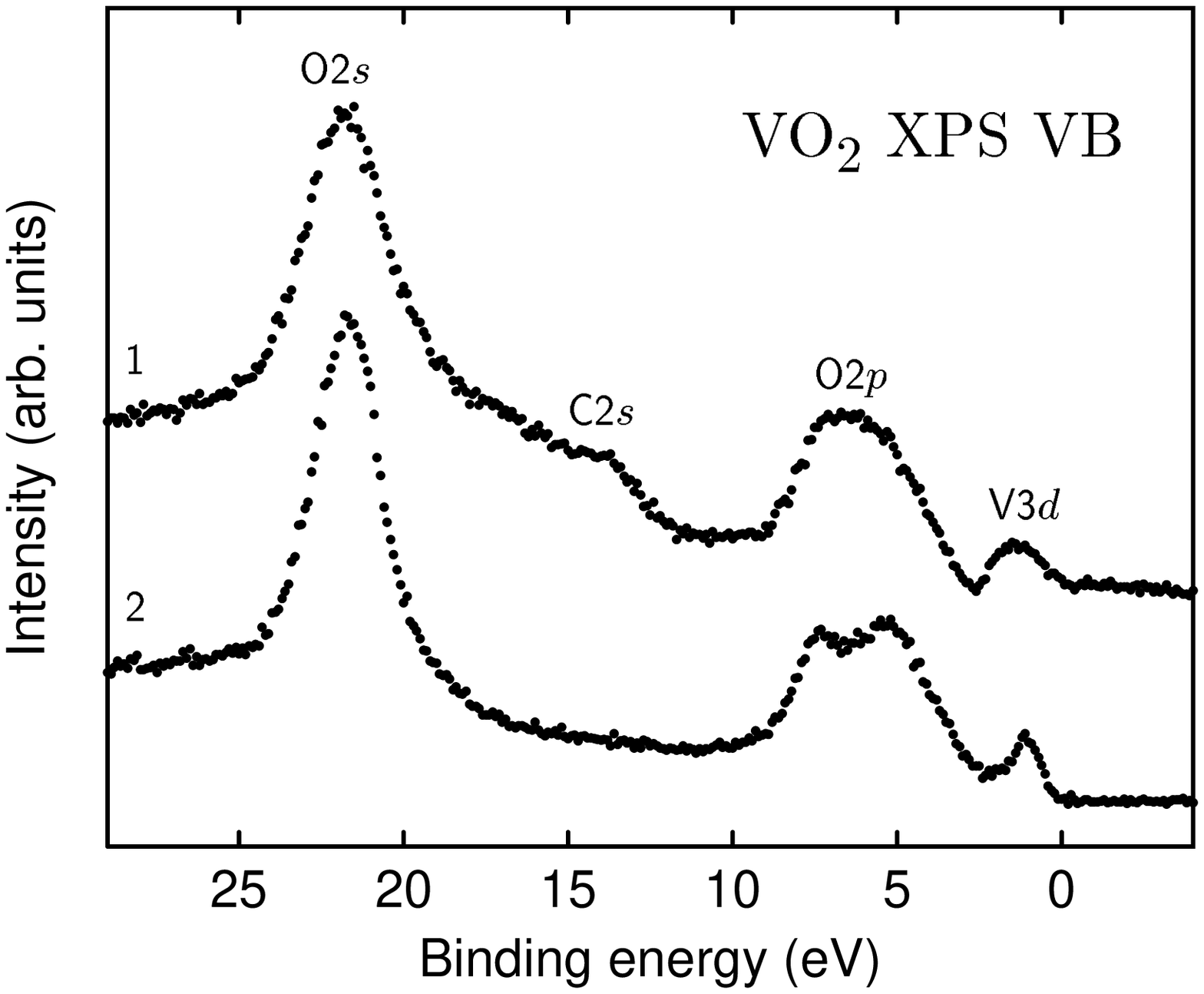}}
\caption{
XPS VB spectrum of uncleaved (1) and cleaved (2)
surface of VO$_2$ single crystal.}
\label{fig_3}
\end{figure}

The intensity distribution of the VB XPS spectra reflects the total DOS of
the VB, up to the deviations due to different atomic photoionization
cross-sections. The results of the XPS measurements of single crystal
VO$_2$ are shown in Fig.~\ref{fig_3} and in the Table 1.

The spectral measurements on the uncleaved crystal show a large contamination
with carbon. This leads to a smearing of the fine structure in the entire
valence band, a broadening of subbands and the appearance of an additional
subband around 12--14 eV which we attribute to transitions from C$2s$ states.

In the XPS spectra of the cleaved VO$_2$ crystal, a distinct narrow peak
is observed around 1 eV below the Fermi level, which, according to the
results of our band structure calculations (see Fig.~\ref{fig_2}), has V$3d$
character. The FWHM of this peak is about 1 eV which is less than obtained
in Ref.~\cite{13,14,15,16}.

\Table{XPS binding energies and width of the core levels (bands) of VO$_2$.}
\br
Core level (band) & Binding Energy (eV) & FWHM (eV) \\
\mr
V$2s$       &  630.02 & 6.10      \\
O$1s$       &  529.75 & 1.30      \\
V$2p_{1/2}$ &  523.48 & 2.56      \\
V$2p_{3/2}$ &  515.95 & 2.04      \\
V$3s$       & \068.95 & 4.54      \\
V$3p$       & \040.53 & 4.24      \\
O$2s$       & \021.73 & 2.5       \\
O$2p$       & \mbox{5.34;~7.24} & \\
V$3d$       & \0\01.03 &          \\
\br
\endtab

The next subband has a two-peak structure at 3--9 eV below the Fermi level
which are mainly due to the O$2p$ states ({\it cf.} Fig.~\ref{fig_2}).
Another reason for reaching this conclusion
is that the energy difference between
the centre of gravity of this band and the next one (located at 22 eV) is
about 15 eV, which is in good agreement with the energy separation
between O$2p$ and O$2s$ bands found in all vanadium oxides \cite{17}.
However, since the O$2p$ and V$3d$ atomic photoionization cross-sections
for Al$K\alpha$ excitation~\cite{11} have a ratio of about 1:2,
one may expect some contribution of V$3d$
states in this band. This is also confirmed by our band structure
calculation and by the results from the V$L\alpha$-emission measurements
(see below).  We point out, however, that there is a contradiction
between the XPS and the ultraviolet photoelectron spectra (UPS) of
VO$_2$ given in Refs.~\cite{13,14,15,16,pes,phB} with respect to
energy resolution and the intensity ratio of the two peaks.
The peaks are well-resolved in our measurements, and
both their energy separation and the intensity ratio are in good agreement
with results of our band structure calculations.

According to Ref.~\cite{18}, charge-transfer satellites are found in the
UPS of ScF$_3$, TiO$_2$ and V$_2$O$_5$ in the region from 11 to 17 eV below
the top of the valence band. From the resonant photoemission measurements
it was concluded that the charge-transfer-type configuration manifests itself
in such a way as to form a valence state in the ground state in addition to
the originally filled $2p$ state ligand. On the basis of this experimental
evidence, cluster calculations of the valence photoemission and
Bremsstrahlung isochromat spectra of VO$_2$ were performed
in Ref.~\cite{6}, and it was concluded that VO$_2$ belongs
to the group of charge-transfer insulators.
However, we point out that our XPS measurements of the uncleaved and cleaved
VO$_2$ single crystal (Fig.~\ref{fig_3}) strongly suggest
that the appearance of this structure (14-15 eV below the Fermi level) is
connected with carbon contamination and that its origin is C$2s$
states. A similar problem was discussed in Ref.~\cite{19} in
connection with the analysis of UPS spectra of superconducting cuprates.
There, it was concluded that the satellites
with binding energies of about 10 eV
are generically connected with contaminations of light elements and
disappear after cleaning.  Therefore we have found a clear indication that
the structure in question is not an intrinsic feature of the electronic
structure of pure VO$_2$ and cannot be considered as an evidence of
correlation effects in VO$_2$.

\begin{figure}[tbh]
\epsfxsize=11.0cm
\centerline{\epsfbox{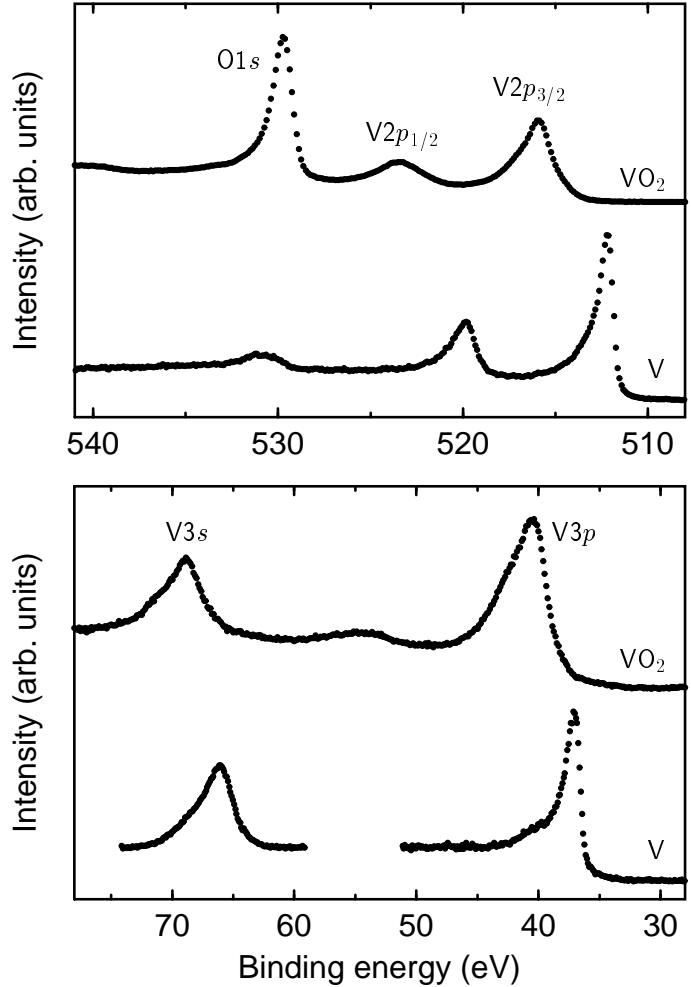}}
\caption{
XPS O$1s$, V$2p_{3/2,1/2}$ (upper panel)
and V$3s$, V$3p$ (lower panel) core level spectra of VO$_2$ and metal V.}
\label{fig_4}
\end{figure}

The XPS core level spectra of VO$_2$ are shown in Fig.~\ref{fig_4}.  The
energy positions of the O$1s$, V$2p$ and V$3p$ spectra of VO$_2$ are close
to those of Ref.~\cite{14}, but the FWHM's of the lines in these spectra
(see Table 1) are much smaller than it was found in Refs.~\cite{13,14}.
The lineshapes are clearly asymmetric due to a structure seen at the
high-binding energy side. In case of $2p$ and $3p$ levels this may be
related to $np^5d^1$ multiplets, as was shown in Ref.~\cite{15}
for the V$3p$ spectrum. It should be noted, however, that in our
measurements the V$3p$ lineshape is somewhat different from that obtained
in Ref.~\cite{15}. In case of V$3s$ level, the additional structure
may be due to the exchange interaction of a (spin-up or
spin-down) 3$s$ electron left in the final state with the
(V$3d$) electrons in the valence band\cite{vV}.

In. Fig.~\ref{fig_4}, the V$2p$, V$3s$ and V$3p$ XPS core level spectra of
VO$_2$ are compared with those of pure vanadium metal. Evidently, there is
a chemical shift of the core level spectra of VO$_2$ with respect to that
of pure metal not only for the V$2p$ XPS-line, but also for the V$3s$
and V$3p$ XPS-lines. The fact that the formal charge of the vanadium atom
affects V$3s$ and V$3p$ lines in an uniform way can be
used for an estimation of the oxidation state of
the V atom in compounds. Recently, the same conclusion was found for the
Mn$3s$ XPS-spectra of manganese complexes in Ref.~\cite{28}.

Earlier, we have shown that the analysis of $3s$ XPS-spectra can be used to
draw conclusions about the electronic structure of $3d$ metal
oxides \cite{22}. It has been found that $3s$ XPS-spectra of NiO have a very
complicated fine structure due to charge-transfer processes.  In this case,
an electron may be transferred from the ligand to a metal $3d$-level after
the emission process, and both states, the screened and unscreened one,
are visible in the $3s$ spectrum. The simpler fine structure of the
V$3s$ XPS-spectrum in VO$_2$ can be considered as an
evidence for negligible charge transfer, again indicating that
the electronic structure of this compound is more bandlike than correlated.

\subsection{X-ray Emission Spectra}
\label{sec:xes}

X-ray emission valence spectra result from electron transitions between
the valence-band and a core hole.  Since the wave function of a core
electron is strongly localized and its angular momentum symmetry is well
defined, these spectra reflect the site-projected and symmetry-restricted
(in accordance with the dipole selection rules) partial DOS.  In the case
of VO$_2$, we have investigated V$L\alpha$- ($3d4s\rightarrow2p _{3/2}$
transition), VK$\beta_5$- ($4p\rightarrow1s$ transition) and
O$K\alpha$-emission spectra ($2p\rightarrow1s$ transition) which reflect
the distribution of V$3d4s$, V$4p$ and O$2p$ partial DOS.  By XPS we can
measure the binding energies of the V$2p$ and the O$1s$ core levels so that
we can determine the position of the Fermi level in the x-ray emission
spectra.

\begin{figure}[tbh]
\epsfxsize=12.0cm
\centerline{\epsfbox{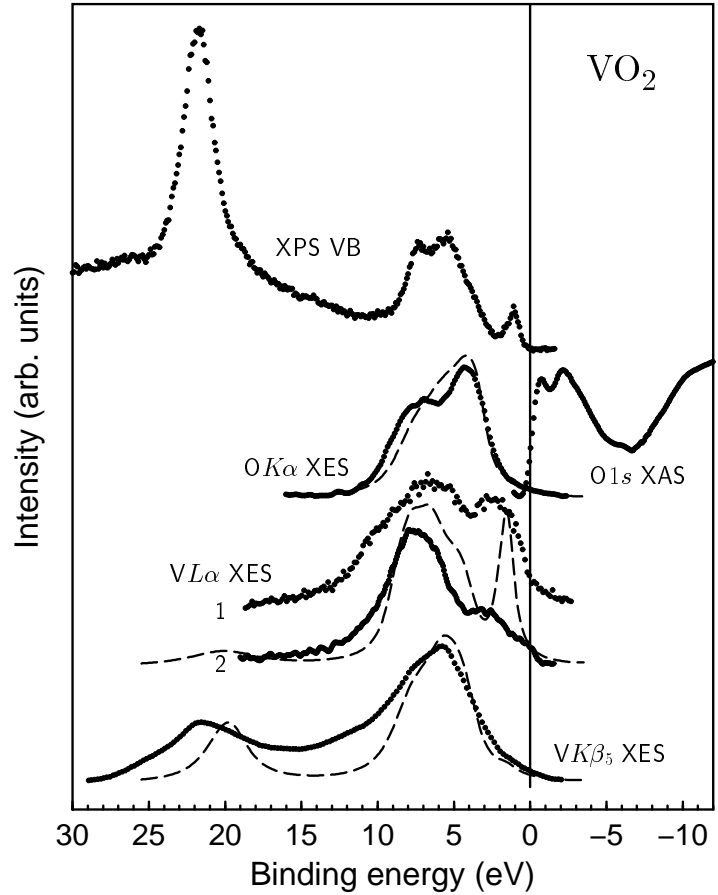}}
\caption{
Dots:
The comparison of XPS VB, XES (OK$\alpha$ (fluorescent excitation at
$E$= 530.8 eV) , V$L\alpha$ (curve 1 corresponds to electron excitation at
$E$=4.4 keV, curve 2 - to fluorescent excitation at $E$=519.0 eV),
V$K\beta_5$ (fluorescent excitation at $E$=15 keV))
and XAS O$1s$ spectra of the VO$_2$ single crystal.
Dashed lines:
broadened partial O$2p$, V$3d$ and V$4p$ (from top to bottom)
DOS as calculated for monoclinic VO$_2$.}
\label{fig_5}
\end{figure}

The O$K\alpha$ and V$L\alpha$ spectra, adjusted in such a way as
to have a a common energy scale with the X-ray photoemission spectrum,
are shown in Fig.~\ref{fig_5}. The
V$K\beta_5$ spectrum is positioned with respect to XPS by matching
the position of the peak which results from the hybridization
with O$2s$ states. In the same figure, the partial DOS
as calculated for the monoclinic phase and broadened with an
energy-dependent lorentzian linewidth according to Ref.~\cite{bs} are shown.
Calculated and broadened partial DOS have been rigidly shifted
with respect to the Fermi energy (arbitrarily, but by the same value
for all spectra) to account for a systematic error in the energy
matching of the measured spectra and calculated DOS due to, e.g.,
the absense of the band gap in the calculation.

As is seen, the main maximum of the O$K\alpha$ spectrum
(O$2p$ states) is placed near the top of the valence band,
that is quite common in oxides, and is essentially nonbonding.
A corresponding feature is also seen in the XPS but merely as
a shoulder at 4 eV, due to a smaller value
of O$2p$ photoionization cross section as compared to V$3d$ \cite{11}.
XPS VB fine sructure with two peaks at 5.4 and 7.5 eV reveals
the O$2p$--V$3d$ bonding band, which overlaps in energy with
the maximum of the V$L\alpha$ spectrum and with a high-binding energy
hump of the O$K\alpha$ spectrum.

We have excited O$K\alpha$ emission at two excitation energies
($E$=530.8 and 532.2 eV) which correspond to the maxima of O$1s$
x-ray absorption spectrum (XAS), but found only little difference.
The V$L\alpha$-emission spectrum has been obtained with
both electron and photon excitation, that resulted in different
intensity distributions (curves 1 and 2 in Fig.~\ref{fig_5}).
With electron excitation at $E$=4.4 keV (curve 1), we have
simultaneous excitation of V$L\alpha$ ($3d4s\rightarrow2p_{3/2}$) and
V$L\beta$-emission ($3d4s\rightarrow2p_{1/2}$) which are separated by only
7.53 eV. Therefore there is an overlap of the high-energy subband
of V$L\alpha$ XES with the main peak of the V$L\beta$-emission spectrum.
For an excitation energy $E$=519.0 eV (curve 2),
we selectively excite V$L\alpha$ which follows electron transitions from the
occupied part of the V$3d$ band. From these data one can conclude that
there is a considerable admixture of V$3d$ states in the O$2p$-like bands,
so that the maximum emission comes from the V$3d$--O$2p$ hybridized band
and not from an only slightly populated V$3d$ band. We find that the fine
structure and energy position of the subbands of the V$L\alpha$-emission
spectrum are in reasonable agreement with the calculated V$3d$ partial DOS
distribution given in Fig.~\ref{fig_2}.

The V$K\beta_5$-emission spectrum has two main subbands whose energy
positions are very close to the O$2p$ and the O$2s$ bands due to
V$4p$--O$2p$ and
V$4p$--O$2s$ hybridization (see Fig.~\ref{fig_2}).  The splitting of the
main subband of the V$K\beta_5$-emission spectrum also follows the
calculated V$4p$ DOS distribution in this energy region. It is seen that
the admixture of the V$4p$ states to the V$3d$ band is too small
to be detected in the V$K\beta_5$ XES of VO$_2$.
The disagreement by $\sim$2 eV between spectra (XPS; V$K\beta_5$)
and calculated DOS in what regards the position of the O$2s$ states
is known in oxides. It is related to the fact that the hole relaxation,
which effectively increases the binding energy of an electron
leaving a comparatively localized state such as O$2s$,
is not taken into account in our band structure calculations which describe
the ground state but not excitations.

It should be mentioned that soft x-ray fluorescence spectra of VO$_2$
(V$L\alpha,\beta$, O$K\alpha$ XES) were also recently reported in
Ref.~\cite{16} and are compared with an UPS spectrum (measured at $E$=501.1
eV): however, no XPS measurements of the V$2p$ and the O$1s$ binding
energies are included there. The O$K\alpha$ XES of Ref.~\cite{16}
does not show the two-peak structure found in our measurements.

\subsection{Total-electron-yield Spectra}

O$1s$ XAS of VO$_2$ probes the O$2p$ unoccupied states.
Such spectra, with good energy resolution, have been published earlier,
by Abbate {\it et al.} \cite{26}. In order to compare the XES
obtained on the same samples which were used in the XPS, we include
in Fig.~\ref{fig_5} our own O$1s$ spectrum obtained (for
the monoclinic phase) in the sample drain-current mode.
The Fermi level has been determined with the help of the XPS
binding energy of the O$1s$ level given in table 1. The
O$1s$ spectrum was measured in a rather restricted energy range, so that we
can only compare the spectra in the vicinity of the Fermi level.

The fine structure of the O$1s$ spectrum is in agreement with the shape
of the O$2p$ conduction band in our electronic structure calculations
(Fig.~\ref{fig_2}). In the high-temperature rutile phase,
the calculated conduction band exhibits a pronounced two-peak structure,
also discussed in Ref.~\cite{26}. In the low-temperature monoclinic phase,
a further splitting of the conduction band is seen from the calculations,
in agreement with a more pronounced structure in the O$1s$ spectrum
of Ref.~\cite{26}.

\section{Conclusion}

The results of measurements of high-resolution x-ray photoelectron spectra,
V$L\alpha$, V$K\beta_5$ and O$K\alpha$ x-ray emission spectra (obtained by
using both electron and x-ray excitation) and O$1s$-absorption
spectra of a VO$_2$ single crystal are presented.  They are compared with
first-principles LMTO band structure calculations of VO$_2$ in monoclinic
and tetragonal rutile phases. It is concluded that the electronic
structure of VO$_2$ is more bandlike than correlated.

\section{Acknowledgements}
Financial support by the Deutsche Forschungsgemeinschaft (SFB~225), the
NATO (grant No.  HTECH.LG971222), and the Russian Foundation for
Fundamental Research (projects No 96-03-32092 and 96-15-96598)
is gratefully acknowledged.  One of us (E.Z.K) wants
to thank the University of Osnabr\"uck for generous hospitality during
his stay.  The Uppsala group gratefully acknowledges financial support by
the Swedish Natural Science Research Council (NFR) and The G\"oran
Gustafsson Foundation and we are indebted to HASYLAB-DESY for the excellent
facilities put to our disposal.

\end{document}